\documentclass[sigconf,nonacm,10pt]{acmart}
\usepackage{tabularx}
\usepackage{booktabs}
\usepackage{tikz}
\usepackage{pgf-pie}
\usepackage{pgfplots}
\usepgfplotslibrary{groupplots}
\usepackage{graphicx}
\usepackage[strings]{underscore}
\usepackage{seqsplit}
\usepackage{array}
\usepackage{xurl}
\usepackage{dblfloatfix}

\newcommand{\code}[1]{\texttt{\seqsplit{#1}}}

\usetikzlibrary{arrows.meta,positioning,fit,backgrounds}
\pgfplotsset{compat=1.18}

\newcolumntype{C}{>{\centering\arraybackslash}X}

\renewcommand\footnotetextcopyrightpermission[1]{}

\begin{document}


\title{PRO-RAN: Processor-Level Characterization of Open RAN Centralized and Distributed Units}

\author{
Moojan Kamalzadeh, Larry Horner, Linqi Xiao, Abhishek Bhattacharyya,\\
Ehsan Bahaloo Horeh, Padmapriya Patil, Venkateswarlu Gudepu, and Andrea Fumagalli
}

\affiliation{
  \institution{The University of Texas at Dallas}
  \city{Richardson}
  \state{Texas}
  \country{USA}
}

\email{mozhan.kamalzadeh@utdallas.edu}

\renewcommand{\shortauthors}{Moojan Kamalzadeh et al.}

\begin{abstract}

Open Radio Access Network (O-RAN) disaggregates RAN protocol functions and enables Centralized Unit (CU) and Distributed Unit (DU) software to execute on general-purpose computing platforms.
Different CU and DU protocol responsibilities produce different processor workloads and execution paths.
Conventional performance metrics, including CPU utilization and throughput, quantify aggregate resource usage without identifying function-level execution costs or processor microarchitectural bottlenecks.
Processor-level characterization, on the other hand, provides insights into resource provisioning, function placement, software optimization, and hardware acceleration.

The paper describes a controlled characterization framework that evaluates independently deployed CU and DU functions under matched hardware and traffic conditions.
The experimental platform integrates the Linux Foundation OCUDU implementation with an emulated User Equipment, a ZeroMQ-based radio interface, and an Open5GS core.
Automated validation confirms registration and bidirectional packet delivery before process-scoped Intel VTune Hotspots and Top-Down Microarchitecture Analysis.
Under traffic load, accumulated process CPU time increases from $17.3$ to $37.4$\,s for the CU and from $462.0$ to $628.4$\,s for the DU during equal 300-s profiling intervals.
The measurements identify distinct CU and DU execution characteristics and motivate function-specific processor analysis and optimization.

\end{abstract}
\maketitle

\section{Introduction}
\label{sec:intro}

Fifth-generation (5G) mobile networks support a wide range of services --- enhanced mobile broadband, industrial automation, autonomous systems, and massive machine-type communication.
Growing throughput, latency, reliability, and scalability requirements increase the computational demands on the Radio Access Network (RAN).
Advances in wireless transmission address radio-side requirements, while programmable computing platforms execute increasingly complex RAN protocol functions.
Software-defined RAN architectures therefore require a detailed understanding of computational behavior for resource provisioning, function placement, and system optimization.


Traditional RAN deployments rely on vertically integrated systems in which a single vendor tightly couples hardware, protocol software, and management components.
Vertical integration limits independent development, replacement, evaluation, and optimization of individual RAN functions.
Open RAN architectures, including those standardized by the O-RAN Alliance, address vertical-integration limitations through functional disaggregation and open standard interfaces between RAN components~\cite{oran-alliance}.
Functional disaggregation enables independent software evolution, multi-vendor interoperability, controlled experimentation, and deployment on general-purpose computing infrastructure.

RAN disaggregation separates protocol processing between the Centralized Unit (CU) and Distributed Unit (DU).
The CU executes higher-layer control-plane and user-plane protocol functions, whereas the DU executes lower-layer radio processing, scheduling, and time-sensitive operations associated with radio transmission.
Different protocol responsibilities produce different computational workloads even when the CU and DU execute on identical processor platforms.
Higher-layer packet processing, control procedures, scheduling, synchronization, and physical-layer processing exercise processor resources through different execution paths and resource-access patterns.


Traditional system-level measurements provide limited visibility into processor-level differences between CU and DU execution.
CPU utilization, throughput, latency, and memory consumption quantify overall resource usage and system performance, but provide limited information about processor-resource consumption.
Two RAN functions may report similar CPU utilization while processor cycles concentrate in different software routines, instruction paths, synchronization operations, cache accesses, or memory-access patterns.
CPU utilization therefore quantifies processor occupancy but does not explain the underlying execution behavior or processor bottlenecks.

Processor-level characterization provides deeper visibility into software-based RAN execution on shared and cloud computing infrastructure.
Function-level profiling identifies software routines responsible for processor-time consumption, while microarchitectural analysis identifies pipeline behavior associated with instruction execution, front-end limitations, back-end limitations, and speculative execution. Detailed characterization supports CPU-core allocation, CU/DU co-location, workload isolation, cloud placement, software optimization, and hardware-acceleration decisions.
Existing experimental studies of open and virtualized RAN platforms predominantly evaluate interoperability, deployment characteristics, throughput, latency, energy consumption, and aggregate CPU utilization~\cite{ferguson2025campus5g,villa2025x5g,bonati2023openrangym,barbosa2025opensource5gcore,guemdani2025comparative,crespo2025energy,wang2020vran}.
Existing measurements provide valuable end-to-end system-level characterization but offer limited processor-level comparison of independently executing CU and DU software functions.

The study presented in this paper focuses on evaluating the processor-level behavior of CU and DU functions under identical hardware and traffic conditions.
The experimental evaluation uses the Linux Foundation-hosted Open Centralized Unit Distributed Unit (OCUDU) implementation as an open-source CU/DU platform~\cite{ocudu2026}.
The experimental platform integrates OCUDU with an emulated User Equipment (UE), a ZeroMQ (ZMQ)-based radio interface, and an Open5GS 5G Core~\cite{open5gs}.

The evaluation characterizes CU and DU processes independently under matched baseline and sustained bidirectional traffic conditions using Intel VTune Profiler.
Hotspots analysis attributes processor time to individual software functions.
Top-Down Microarchitecture Analysis (TMA)~\cite{yasin2014topdown} categorizes processor pipeline slots into Retiring, Front-End Bound, Back-End Bound, and Bad Speculation.
Joint Hotspots and TMA analysis links software-level execution paths to processor microarchitectural behavior.
The combined analysis identifies both \emph{where} CU and DU processes consume processor time and \emph{how} CU and DU execution utilizes processor resources.

The key contributions of this study are as follows:

\begin{itemize}
\item A reproducible end-to-end methodology for processor-level characterization of independently deployed 5G CU and DU functions using automated network validation and process-scoped Intel VTune profiling.

\item A controlled CU/DU comparison on identical computing platforms under matched baseline and sustained bidirectional-load conditions.

\item A function-level Hotspots analysis combined with TMA to relate software execution paths to processor microarchitectural behavior.

\item A processor-level characterization that identifies distinct CU and DU execution behavior beyond aggregate CPU-utilization measurements, supporting function-aware provisioning, placement, optimization, and hardware acceleration.

\item An Ansible-based automation platform for deployment, traffic generation, end-to-end validation, profiling, symbol resolution, and result collection (outside the scope of the paper).

\end{itemize}
\section{Related Work}
\label{sec:related}

Open-source O-RAN platforms support deployment, experimentation, and interoperability evaluation.
Campus5G provides a campus-scale private 5G Open RAN platform~\cite{ferguson2025campus5g}, X5G provides a programmable multi-vendor 5G platform with GPU-accelerated Physical-layer (PHY) processing and a near-real-time RAN Intelligent Controller~\cite{villa2025x5g}, and OpenRAN Gym provides an Artificial Intelligence / Machine Learning (AI/ML) framework for data collection and experimentation on O-RAN testbeds~\cite{bonati2023openrangym}.
System-level studies further evaluate open-source 5G implementations using throughput, latency, scalability, computational efficiency, and aggregate CPU utilization~\cite{barbosa2025opensource5gcore,guemdani2025comparative}.
Platform-level and system-level evaluations characterize deployment and end-to-end performance without function-level attribution of processor execution across independently deployed CU and DU processes.

Processor-level studies analyze software execution using hardware performance counters and microarchitectural metrics.
The work in~\cite{crespo2025energy} profiles an srsRAN DU using Linux \texttt{perf} counters --- CPU utilization, cache misses, context switches, and Instructions Per Cycle (IPC) --- for CPU-affinity and frequency selection.
Another work~\cite{wang2020vran} applies Intel VTune to a monolithic OAI gNB/UE platform co-located with Multi-access Edge Computing applications and reports thread-level CPU usage and top-down microarchitectural metrics.
Existing processor-level studies characterize a DU implementation or a monolithic RAN configuration without a matched comparison between independently deployed CU and DU processes.
As previously stated, the proposed work characterizes Linux Foundation OCUDU CU and DU processes under matched baseline and sustained-load conditions using process-scoped Hotspots analysis and TMA, enabling direct comparison of function-level CPU-time distribution and processor microarchitectural behavior.

\begin{figure}[!htbp]
\centering
\includegraphics[
  width=\linewidth,
  trim=0cm 3cm 0cm 0.5cm,
  clip
]{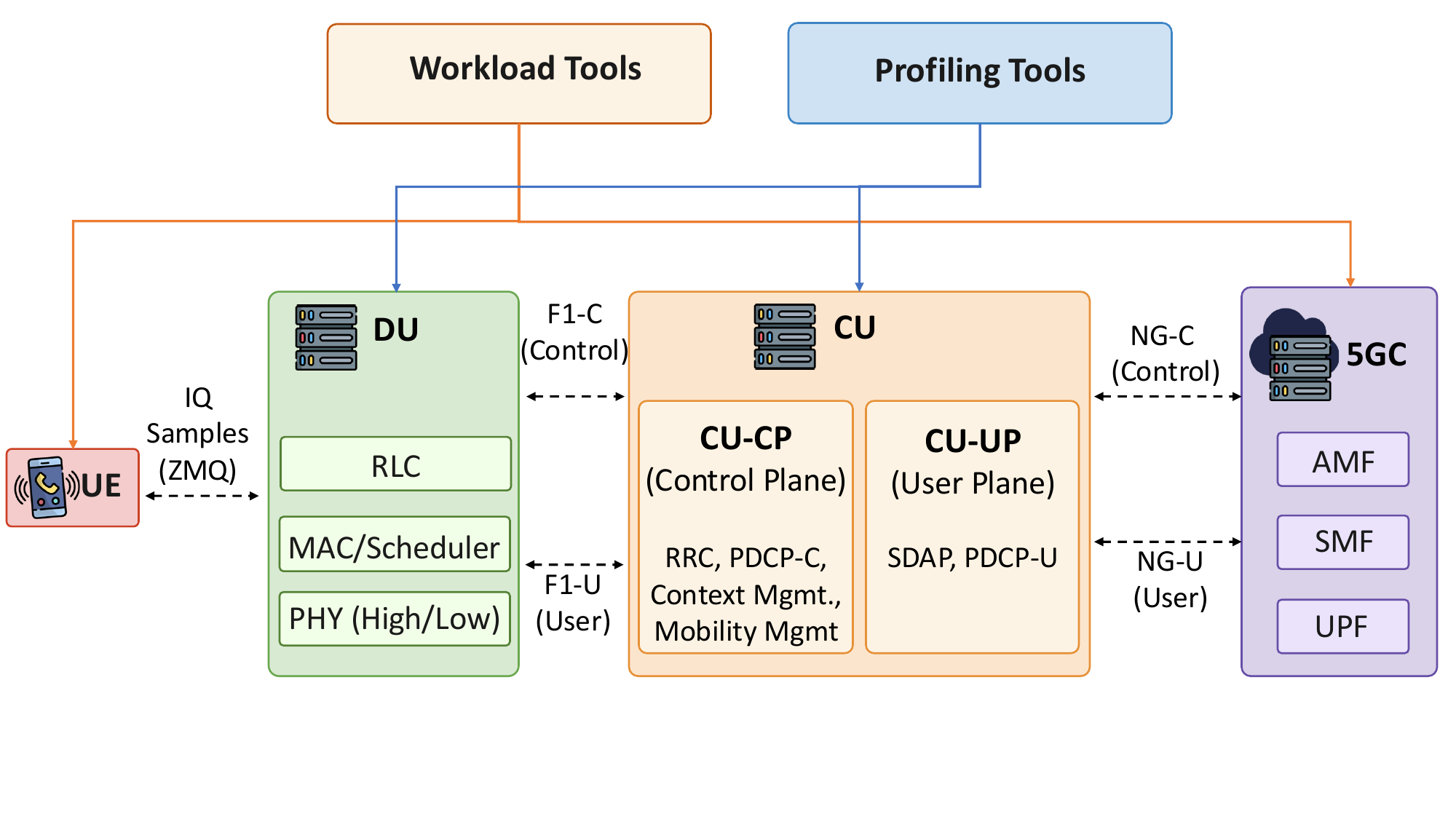}
\caption{End-to-end 5G system model}
\label{fig:system-model}
\end{figure}

\section{System Model}
\label{sec:system-model}

The system model consists of an end-to-end 5G architecture and a processor-profiling framework, as shown in Figure~\ref{fig:system-model}.
The 5G architecture includes an emulated UE, DU, CU, and a 5G Core (5GC).
The UE exchanges in-phase and quadrature (IQ) samples with the DU through a ZMQ-based radio interface.
The DU executes lower-layer RAN functions --- PHY processing, Medium Access Control (MAC) scheduling, and Radio Link Control (RLC).
The CU executes higher-layer protocol functions and separates control-plane and user-plane processing.
The CU Control Plane (CU-CP) executes Radio Resource Control (RRC), control-plane Packet Data Convergence Protocol (PDCP), context management, and mobility management, while the CU User Plane (CU-UP) executes Service Data Adaptation Protocol (SDAP) and user-plane PDCP.
F1-C and F1-U interfaces connect the DU with the CU for control-plane and user-plane communication, respectively.
The CU connects with the 5GC through NG-C and NG-U interfaces.
The 5GC includes the Access and Mobility Management Function (AMF), Session Management Function (SMF), and User Plane Function (UPF).

The framework also includes workload-generation and processor-profiling tools.
Workload tools generate controlled end-to-end traffic between the UE and 5GC and establish matched baseline and sustained-load operating conditions.
Profiling tools monitor the independently executing CU and DU processes under each operating condition.
Process-scoped Hotspots analysis quantifies CPU-time distribution across software functions, while TMA identifies processor bottlenecks.
The separation between workload generation and process profiling maintains identical traffic conditions for CU and DU measurements and enables direct comparison of software execution and microarchitectural behavior.

\begin{figure*}[!htb]
\centering
\includegraphics[
  width=\textwidth,
  trim={0.72cm 2cm 0.7cm 3cm},
  clip
]{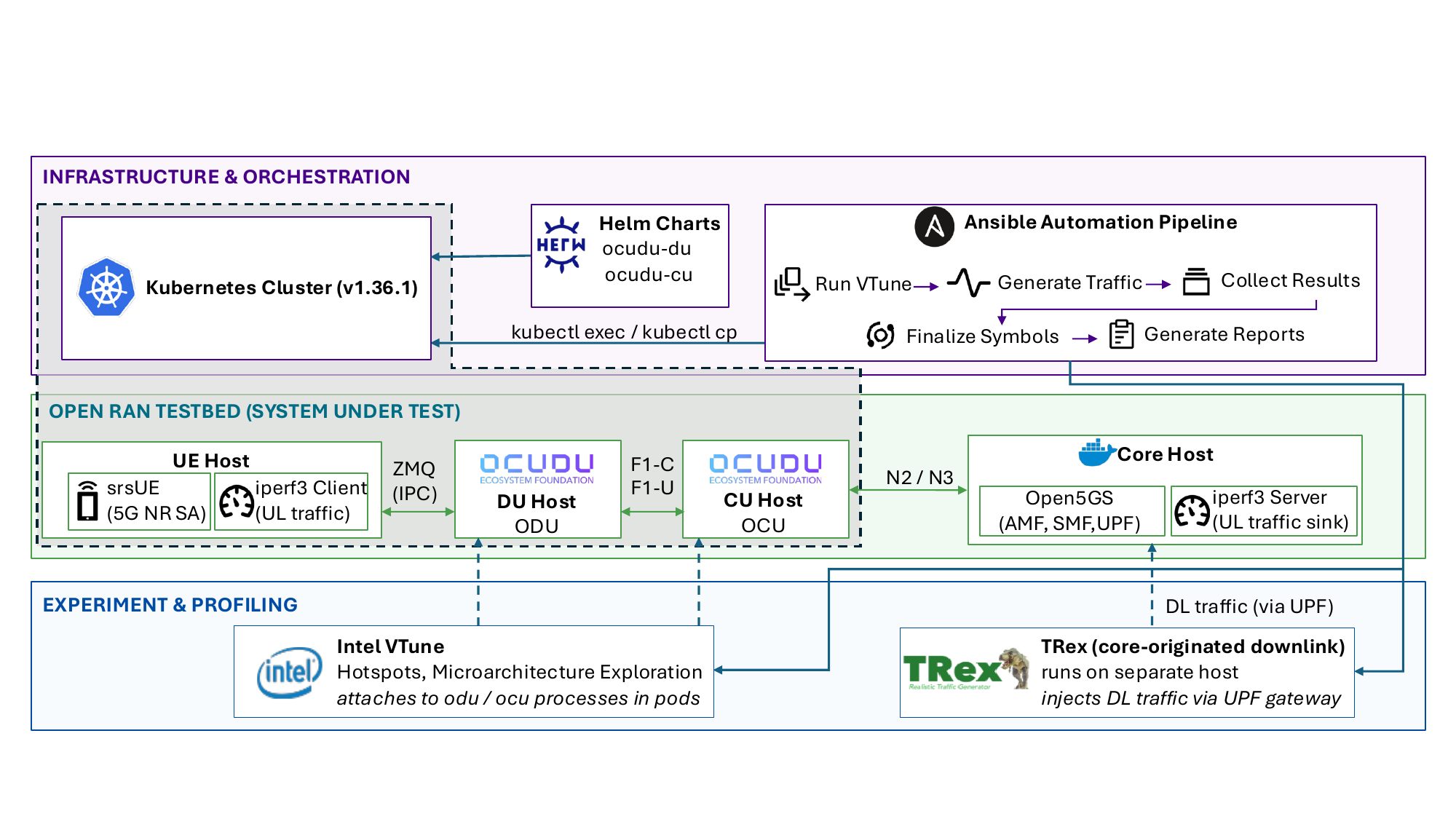}
\caption{Experimental setup of the OCUDU-based 5G testbed}
\label{fig:exp-setup}
\end{figure*}
\section{Experimental Setup}
\label{sec:method}

The experimental setup implements the end-to-end 5G system along with automated profiling tools as depicted in Figure~\ref{fig:exp-setup}.
An end-to-end standalone 5G deployment is built using an emulated UE, an OCUDU DU and CU, and an Open5GS core network. 
The platform supports automated deployment, traffic generation, end-to-end validation, and process-scoped processor profiling.


\subsection{Platform and Deployment}
\label{sec:platform}

Separate hosts execute srsUE, one OCUDU DU, one OCUDU CU, and Open5GS.
The experiments use OCUDU commit \code{ef8af93d}, srsUE commit \code{6bcbd9e}, and Open5GS v2.7.6.
The UE and DU exchange digitized IQ samples through a ZMQ-based radio interface.
Required PHY, ZMQ-RF, and Non-Access Stratum (NAS) modifications enable ZMQ-based cell acquisition and 5G registration.
The UE host also executes the uplink traffic client, while the core-network side hosts the corresponding traffic endpoint.

The CU and DU hosts use identical single-socket processor platforms with 40 physical cores and 80 logical processors. Memory, operating-system, kernel, CPU-frequency, CPU-affinity, Simultaneous Multithreading (SMT), and container resource configurations remain identical across all measurements.
The common configuration controls processor-platform variation during CU/DU comparison.
Specific processor models and server-identifying hardware details are omitted in accordance with the testbed confidentiality agreement.

The OCUDU CU and DU are deployed as independent Kubernetes workloads running the \texttt{ocu} and \texttt{odu} processes, respectively.
An Ansible-based workflow controls deployment, traffic execution, network validation, Intel VTune collection, symbol resolution, and result extraction.

\subsection{Radio and Network Configuration}
\label{sec:network-config}

Experiments use standalone 5G New Radio (NR) with Frequency Division Duplex (FDD) operation in band n3.
A 10\,MHz channel bandwidth and 15\,kHz subcarrier spacing provide 52 Physical Resource Blocks (PRBs).
The Downlink (DL) Absolute Radio Frequency Channel Number (ARFCN) is 368500, and DL and Uplink (UL) transmissions support modulation up to 64-Quadrature Amplitude Modulation (64-QAM).
One emulated UE uses Public Land Mobile Network (PLMN) identifier 00101 and Tracking Area Code (TAC) 7.

The CU and DU use the Third Generation Partnership Project (3GPP) Option~2 functional split, with F1-C and F1-U carrying control- and user-plane traffic, respectively.
A ZMQ-based radio interface exchanges digitized IQ samples between the DU and UE without physical Radio Frequency (RF) hardware.
Radio timing, scheduling, Physical-layer (PHY) processing, UE registration, and Protocol Data Unit (PDU) session procedures remain active during the experiments.
It is worth noting that the OCUDU testmode supports multiple UEs, which will be considered in future work.




\subsection{Traffic Generation and Validation}
\label{sec:traffic}

The evaluation considers baseline and sustained-load operating conditions under identical radio, software, and hardware configurations.
The baseline condition maintains UE registration and an established PDU session without application-layer offered traffic. 
The load condition applies sustained bidirectional traffic using fixed DL and UL traffic parameters.

DL User Datagram Protocol (UDP) traffic originates from TRex~\cite{trex} and traverses the 5GC user plane toward the UE.
UL traffic originates from an \texttt{iperf3} client bound to the UE tunnel interface and terminates at an \texttt{iperf3} server on the core-network side.

The automation workflow validates end-to-end operation before every profiling collection.
Validation confirms UE registration, PDU-session establishment, and bidirectional packet delivery.
DL validation verifies CU PDCP activity and UE Physical Downlink Shared Channel (PDSCH) reception.
UL validation verifies DU Buffer Status Report (BSR) activity, Physical Uplink Shared Channel (PUSCH) transmission, CU PDCP activity, and core-side packet reception.
Processor profiling starts after successful validation and stable traffic delivery.




\subsection{Profiling and Symbol Resolution}
\label{sec:symbols}

Intel VTune Profiler 2025.4~\cite{vtune} attaches to the process identifiers of the \texttt{ocu} and \texttt{odu} processes for process-scoped profiling.
Hotspots and Microarchitecture Exploration collections execute independently for CU and DU under baseline and sustained-load conditions, producing eight 300-s collections across two processes, two traffic conditions, and two profiling analyses.
OCUDU executables and supporting runtime, ZMQ, 
and kernel components retain debug symbols for function-level attribution.

Hotspots analysis quantifies CPU time across software functions and libraries, while TMA classifies pipeline slots into Retiring, Front-End Bound, Back-End Bound, and Bad Speculation.
{\em Retiring} represents completed instructions, {\em Front-End Bound} represents instruction-delivery limitations, {\em Bad Speculation} represents discarded work, and {\em Back-End Bound} represents unavailable data or execution resources.
\begin{figure}[!htb]
\centering
\includegraphics[
  width=\linewidth,
  trim={0cm 0cm 0cm 2cm},
  clip
]{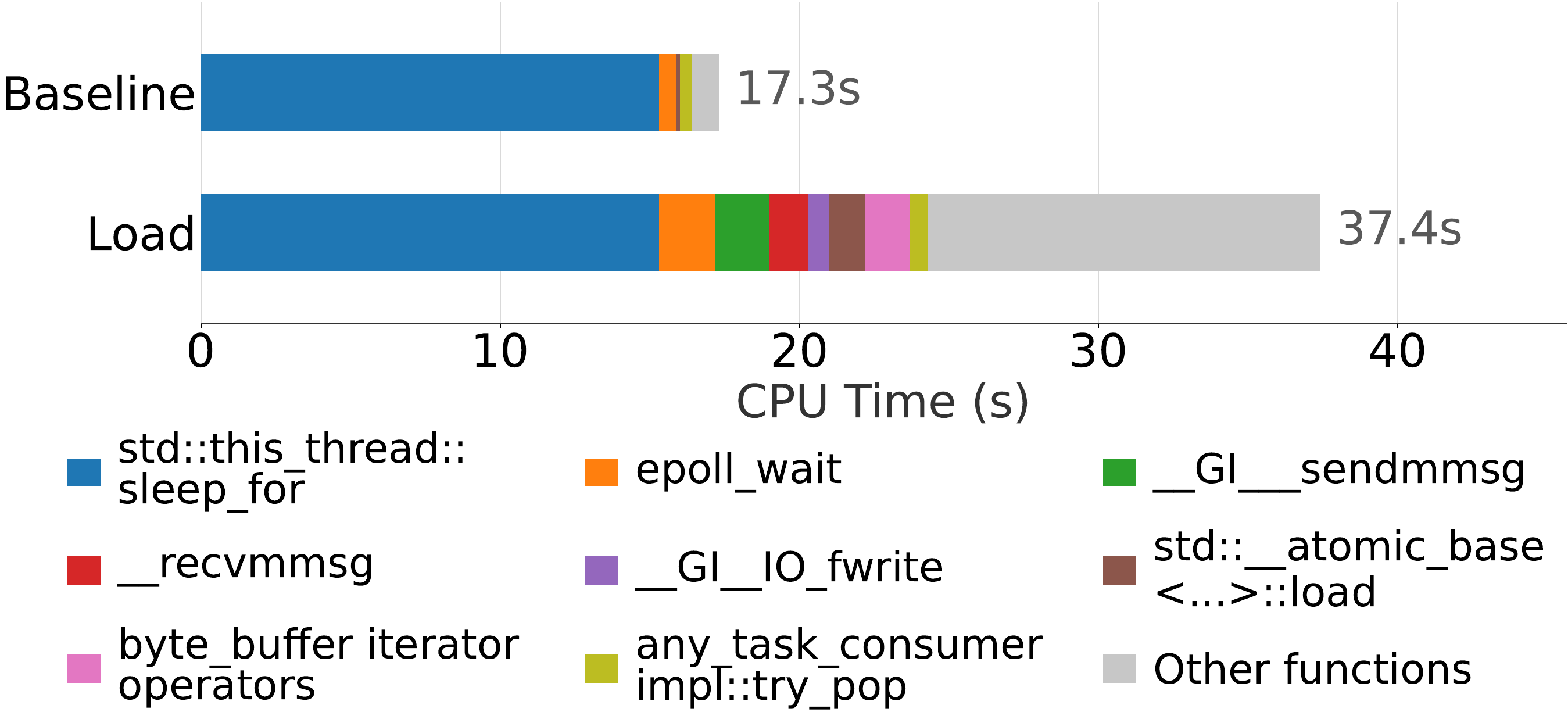}
\caption{Whole-process CU hotspots}
\label{fig:hotspot-composition-cu}
\end{figure}

\begin{figure}[!htb]
\centering
\includegraphics[
  width=\linewidth,
  trim={0cm 0cm 0cm 2cm},
  clip
]{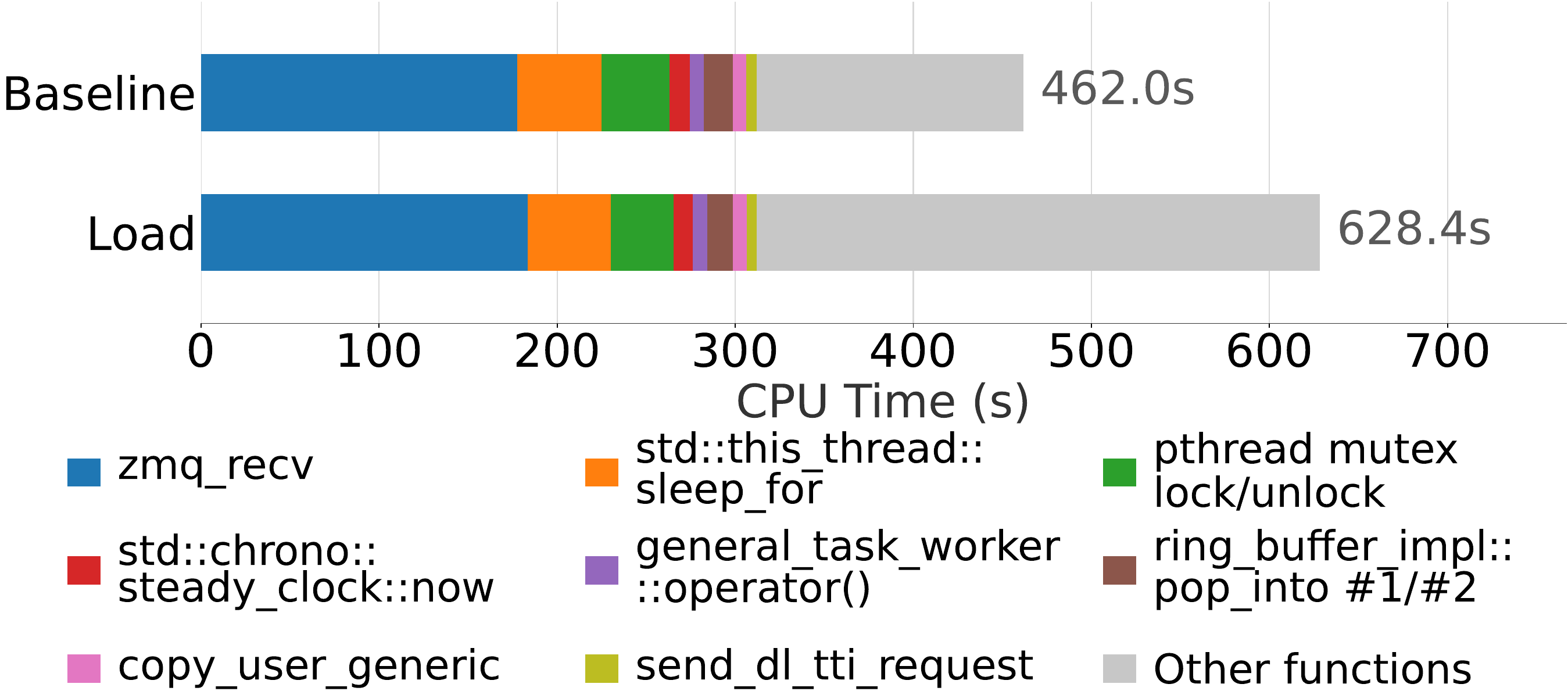}
\caption{Whole-process DU hotspots}
\label{fig:hotspot-composition-du}
\end{figure}

\begin{table*}[!htb]
\centering
\caption{Displayed hotspot functions, software operations, and CPU times}
\label{tab:hotspot-map}
\scriptsize
\setlength{\tabcolsep}{3pt}
\renewcommand{\arraystretch}{1.08}

\begin{tabularx}{\textwidth}{@{}
p{6.5cm}
l
p{2.5cm}
r
r
r
@{}}
\toprule
\textbf{Hotspot function(s)} &
\textbf{Software operation} &
\textbf{Category} &
\textbf{Base. (s)} &
\textbf{Load (s)} &
\textbf{$\Delta$ (s)} \\
\midrule
\multicolumn{6}{c}{\textbf{Centralized Unit (CU)}}\\
\midrule

\code{std::this_thread::sleep_for}
&
Worker sleep between processing intervals
&
Sleep/wait
&
15.3 & 15.3 & 0.0
\\

\code{epoll_wait}
&
Socket-event monitoring
&
libc / system
&
0.6 & 1.9 & +1.3
\\

\code{__GI___sendmmsg}
&
Batch packet transmission
&
libc / system
&
0.0 & 1.8 & +1.8
\\

\code{__recvmmsg}
&
Batch packet reception
&
libc / system
&
0.0 & 1.3 & +1.3
\\

\code{__GI__IO_fwrite}
&
Runtime output
&
libc / system
&
0.0 & 0.7 & +0.7
\\

\code{std::__atomic_base<...>::load}
&
Shared-state access
&
Runtime \& sync
&
0.1 & 1.2 & +1.1
\\

\code{byte_buffer_segment_byte_iterator_impl::operator++}
\newline
\code{byte_buffer_segment_byte_iterator_impl::operator==}
&
Segmented-buffer traversal and iterator comparison
&
OCUDU buffer traversal
&
0.0 & 1.5 & +1.5
\\

\code{any_task_consumer_impl::try_pop}
&
Task retrieval
&
OCUDU task queues
&
0.4 & 0.6 & +0.2
\\

\emph{Other Functions} &
Lower-ranked functions &
Multiple categories &
0.9 & 13.1 & +12.2
\\

\addlinespace[2pt]
\midrule
\multicolumn{6}{c}{\textbf{Distributed Unit (DU)}}\\
\midrule

\code{zmq_recv}
&
IQ-sample reception
&
ZMQ transport
&
177.5 & 183.5 & +6.0
\\

\code{std::this_thread::sleep_for}
&
Worker sleep between processing intervals
&
Sleep/wait
&
47.4 & 46.6 & -0.8
\\

\code{___pthread_mutex_lock}
\newline
\code{___pthread_mutex_unlock}
&
Mutex coordination
&
Runtime \& sync
&
38.1 & 35.2 & -2.9
\\

\code{std::chrono::_V2::steady_clock::now}
&
Timing reference
&
Runtime \& sync
&
11.4 & 11.0 & -0.4
\\

\code{general_task_worker::}
\newline
\code{make_blocking_pop_task::operator()}
&
Blocking task retrieval
&
OCUDU task queues
&
8.1 & 7.9 & -0.2
\\

\code{ring_buffer_impl::pop_into \#1}
\newline
\code{ring_buffer_impl::pop_into \#2}
&
Ring-buffer retrieval
&
OCUDU ring buffers
&
16.2 & 14.6 & -1.6
\\

\code{copy_user_generic}
&
User--kernel memory copy
&
Linux kernel
&
7.4 & 7.6 & +0.2
\\

\code{fapi_to_phy_fastpath_translator::send_dl_tti_request}
&
Downlink scheduling translation
&
Scheduler / PHY--FAPI
&
5.8 & 5.5 & -0.3
\\

\emph{Other Functions} &
Lower-ranked functions &
Multiple categories &
150.1 & 316.5 & +166.4
\\

\bottomrule
\end{tabularx}

\vspace{2pt}
\begin{minipage}{\textwidth}
\footnotesize
Rows containing multiple named functions report the combined CPU time
of the displayed instances.
\end{minipage}
\end{table*}

\vspace{-1pt}
\section{Results}
\label{sec:results}
The evaluation characterizes the execution of CU and DU at the function and microarchitectural levels under baseline and sustained-load conditions.
The results are discussed in the following subsections.

\subsection{Function-Level CPU-Time Distribution}
\label{sec:results-hotspots}

VTune CPU time measures how long the profiled process actively executes on the processor. CPU time is accumulated across all process threads and can therefore exceed the profiling duration when multiple threads execute concurrently.
Figures~\ref{fig:hotspot-composition-cu} and \ref{fig:hotspot-composition-du} show the whole-process CPU-time distribution, while Table~\ref{tab:hotspot-map} maps each displayed function to the corresponding software operation and processing category and reports baseline, load, and CPU-time change. 
Rows containing multiple functions represent combined CPU time, while \emph{Other Functions} aggregates lower-ranked functions below the display threshold.



For the CU, Figure~\ref{fig:hotspot-composition-cu} and
Table~\ref{tab:hotspot-map} show a clear load-dependent shift in execution.
Sleep/wait time remains nearly constant, while packet I/O, buffer traversal,
shared-state access, and task-queue operations increase under sustained
traffic.
Packet transmission and reception also emerge under load.
CU execution therefore shifts from an event-waiting baseline toward active
user-plane packet processing and worker coordination.
The \emph{Other Functions} contribution also increases, indicating additional
processing across multiple lower-ranked runtime, synchronization, and
support functions rather than one dominant hotspot.

For the DU, Figure~\ref{fig:hotspot-composition-du} and Table~\ref{tab:hotspot-map} show a different execution pattern.
IQ-sample reception remains the dominant operation under both conditions, while
timing, synchronization, task-queue, ring-buffer, memory-copy, and scheduling functions remain approximately stable.
Most additional DU CPU time appears in lower-ranked functions.
Continuous IQ processing, radio timing, and scheduling remain active without sustained application traffic, creating a substantial traffic-independent DU processing baseline.
In contrast, the CU maintains UE context and session state during baseline operation and activates substantial user-plane processing under traffic.
The measurements therefore characterize the DU as predominantly timing-driven and the CU as more traffic-responsive.

\subsection{Microarchitecture Behavior}
\label{sec:results-uarch}

\begin{figure}[!htb]
\centering
\includegraphics[
  width=\linewidth,
  trim={0cm 3cm 1cm 2cm},
  clip
]{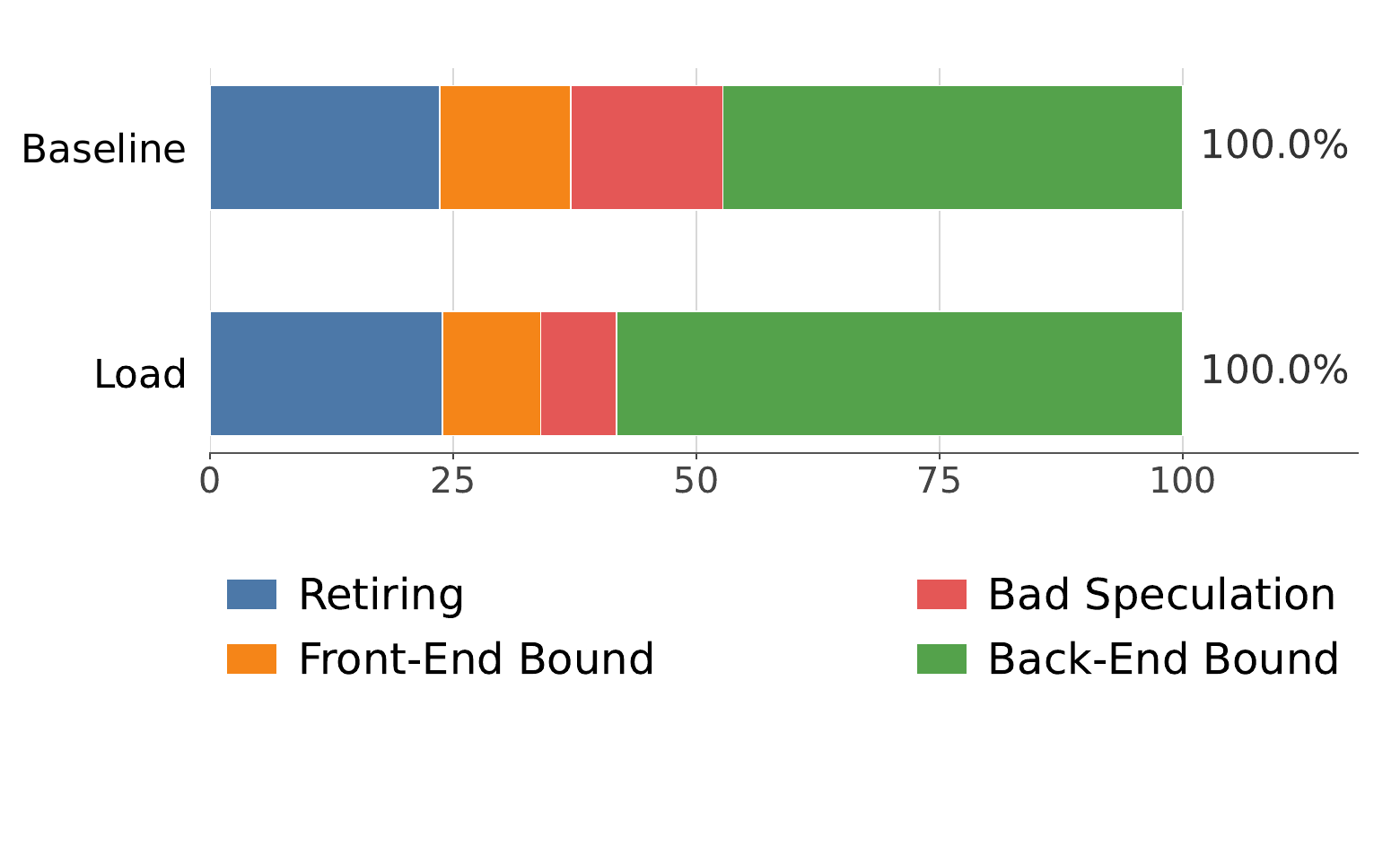}
\caption{Top-level TMA metrics (CU)}
\label{fig:uarch-composition-cu}
\end{figure}

\begin{figure}[!htb]
\centering
\includegraphics[
  width=\linewidth,
  trim={0cm 3cm 1cm 2cm},
  clip
]{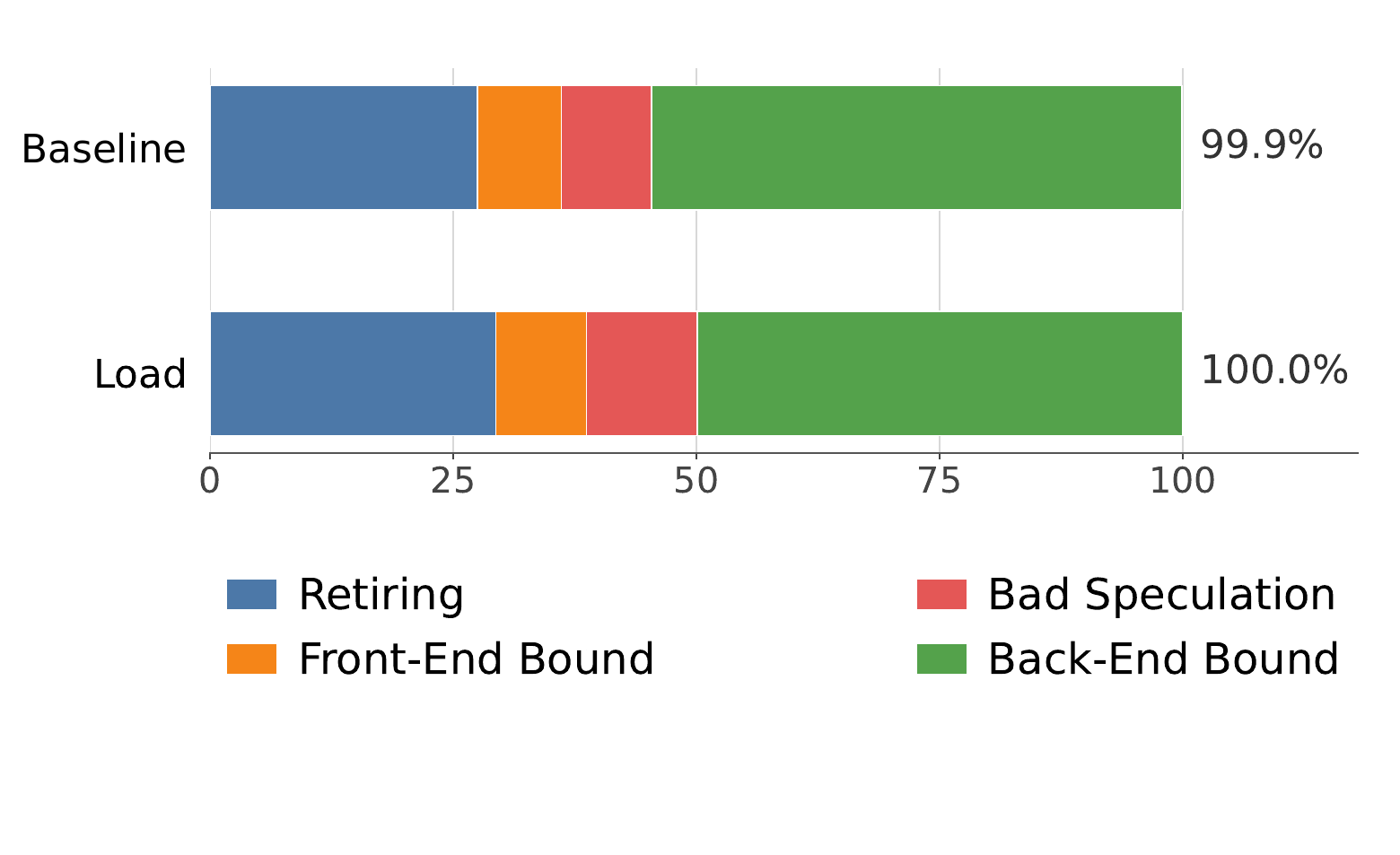}
\caption{Top-level TMA metrics (DU)}
\label{fig:uarch-composition-du}
\end{figure}

Figures~\ref{fig:uarch-composition-cu} and
\ref{fig:uarch-composition-du} show the top-level TMA distributions, while Table~\ref{tab:tma-summary} reports Retiring, Front-End Bound, Bad Speculation, and Back-End Bound under baseline and load conditions.
The TMA categories quantify useful instruction execution, instruction-delivery limitations, discarded speculative work, and back-end resource limitations,
respectively.


For the CU, Figure~\ref{fig:uarch-composition-cu} and
Table~\ref{tab:tma-summary} show a clear load-dependent shift toward back-end limitations.
Back-End Bound increases from 47.3\,\% to
58.2\,\%, while Retiring remains nearly constant at 23.6--23.9\,\%.
Front-End Bound decreases from 13.5\,\% to 10.1\,\%, and Bad Speculation decreases from 15.6\,\% to 7.8\,\%. CU execution therefore experiences greater back-end pressure under sustained traffic without a corresponding increase in the Retiring share.


\begin{table}[t]
\centering
\caption{Top-level TMA metrics (\%)}
\label{tab:tma-summary}
\footnotesize
\begin{tabular*}{\linewidth}{@{\extracolsep{\fill}}lrrrr@{}}
\toprule
Metric & CU Base. & CU load & DU Base. & DU load \\
\midrule
Retiring & 23.6 & 23.9 & 27.5 & 29.4 \\
Front-End Bound & 13.5 & 10.1 & 8.6 & 9.3 \\
Bad Speculation & 15.6 & 7.8 & 9.3 & 11.4 \\
Back-End Bound & 47.3 & 58.2 & 54.5 & 49.9 \\
\bottomrule
\end{tabular*}
\end{table}


For the DU, Figure~\ref{fig:uarch-composition-du} and Table~\ref{tab:tma-summary} show a more stable microarchitectural distribution.
Back-End Bound remains the dominant category despite decreasing from 54.5\,\% to 49.9\,\%, while Retiring increases from 27.5\,\% to 29.4\,\%.
Front-End Bound changes from 8.6\,\% to 9.3\,\%, and Bad Speculation changes from 9.3\,\% to 11.4\,\%.
Continuous radio processing during baseline operation produces a substantial processor workload before sustained application traffic, resulting in a smaller baseline-to-load microarchitectural shift.

Overall, Hotspots and TMA show different CU and DU load responses.
The CU shows a clear processor-level shift, while Retiring remains nearly unchanged.
The DU remains predominantly Back-End Bound under both conditions.
Traffic increases CU activity across packet-processing and coordination functions.
DU load increases CPU time mainly across lower-ranked functions, while leading operations remain stable.
The measurements therefore show distinct CU and DU execution characteristics under matched conditions.
\section{Summary and Future Work}
\label{sec:conclusion}

A processor-level characterization framework is presented for independently deployed OCUDU CU and DU functions in an end-to-end 5G system.
Process-scoped Intel VTune Hotspots and TMA quantify function-level CPU-time distribution and processor microarchitectural behavior under matched baseline and sustained-load conditions.
Experimental results show accumulated process CPU time increasing from $17.3$ to $37.4$\,s for the CU and from $462.0$ to $628.4$\,s for the DU during equal 300-s profiling intervals.
CU execution exhibits a stronger load-dependent microarchitectural shift, with Back-End Bound increasing from 47.3\,\% to 58.2\,\%, while DU behavior changes more gradually.
The measured differences motivate function-specific resource provisioning, placement, software optimization, and hardware acceleration.

Future work focuses on multi-UE workloads, additional traffic profiles, broader radio configurations, and processor-level evaluation across additional CU/DU implementations and computing platforms.




\begin{acks}
The work reported here is supported by NSF grant CNS-1956357.
\end{acks}

\bibliographystyle{ACM-Reference-Format}
\bibliography{references}

\end{document}